\documentclass[12pt]{article}
\begin{document}
\parindent 0 pt


\def \ep{\epsilon} 
\def \intR {\int_{-\infty}^{+\infty}}  
\def \grad {\nabla}  
\def \ov{\over} 
\def \q  {\quad}  
\def \qq {\qquad}  
\def \bar{\overline} 
\def \beq{ \begin{equation} } 
\def \eeq{\end{equation}} 
\def \r#1{$^{#1}$}
\newtheorem{proposition}{Proposition}
\newtheorem{lemma}{Lemma}                 
\newtheorem{theorem}{Theorem}
\renewcommand{\labelenumi}{(\roman{enumi})}


\title{Symmetric Periodic Solutions of the Anisotropic Manev Problem}

\author{{Manuele  Santoprete}\renewcommand{\thefootnote}{\alph{footnote})}\footnotemark[1]
\\ Department of  Mathematics and Statistics \\  University of 
 Victoria, P.O.  Box  3045 Victoria B.C., \\ Canada, V8W 3P4}

\renewcommand{\thefootnote}{\alph{footnote})}

\date{}

\maketitle
\footnotetext[1]{Electronic mail: msantopr@math.uvic.ca}
\begin{abstract}
\noindent We consider  the Manev potential  in an anisotropic space,  i.e., such
that  the   force  acts  differently  in  each   direction.   Using  a
generalization  of the  Poincar\'e  continuation method  we study  the
existence of periodic solutions  for weak anisotropy. In particular we
find  that the  symmetric periodic  orbits  of the  Manev system  are
perturbed to periodic orbits in the anisotropic problem.
\end{abstract}

\vskip 0.5truecm
\hspace{25pt}  PACS(2001): 05.45.-a, 45.50.Jf, 45.50.Pk 

\vfill\eject
~
\vskip 2truecm

\small\normalsize
\section*{\large\bf I. Introduction}

In this paper we consider the Anisotropic Manev Problem (AMP) that was
introduced by Diacu\r{1}  in the early 1990s.  The work on the
AMP  was inspired  by  the Anisotropic  Kepler  Problem introduced  by
Gutzwiller in  the early 1970s.  Gutzwiller aimed to  find connections
between classical  and quantum mechanics. His  interest was stimulated
by an old  unsolved quantum mechanical problem formulated  in a paper
written     by    Einstein:\r{2}     even    if     the
Born-Sommerfeld-Einstein  condition  (e.g.  see Ref. 2)  were
appropriate to  describe the semi-classical limit of  quantum theory it
was unclear  how to find  a classical approximation  for nonintegrable
systems.

Similarly  the main  reason  for  considering the  AMP  is to  further
analyze   similarities  between   classical   mechanics  and   quantum
theory. Moreover,  as it was  remarked in Ref. 1, the  AMP also
brings general relativity into the game, since, the Manev's potential,
explains the  perihelion advance  of the inner  planets with  the same
accuracy as general relativity.\r{3}  It should be remarked
that  bringing  general relativity  into  the  game  is of  particular
importance since a satisfactory quantum theory of gravitation does not
exist.

Some of the qualitative features of the Anisotropic Manev Problem have
already   been  studied.    In Ref. 1,  a   large  class   of
capture-collision and ejection-escape solutions is studied by means of
the  collision and  infinity manifold  techniques. In  particular that
paper   also   brought   arguments   favouring  the   chaoticity   and
nonintegrability   of  the   system  by   showing  the   existence  of
heteroclinic orbits  within the zero energy  manifold. In Ref. 4
the  occurrence  of  chaos  on   the  zero  energy  manifold  and  the
nonintegrability are  finally proved,  putting into evidence  that the
AMP  is a  very  complex problem.  

In  this work,  to  gain  a better  understanding  of the  complicated
dynamics of  the AMP, we find  the symmetric periodic orbits. 
Analyzing those orbits is especially important since,
by now, it  is well known that studying periodic  orbits is a valuable
general approach  to tackle  complex problems in  classical mechanics.
The existence  of periodic  orbits  for small
values of  the anisotropy is proved  using generalizations
of  the  Poincar\'e  continuation  method developed  in Refs. 5-7.

The (planar) anisotropic Manev problem is described by the Hamiltonian
\beq H={1 \over  2}{\bf p}^2 - {1 \over  \sqrt{x^2+ \mu y^2}}-{b \over
x^2 +\mu y^2}.
\label{H}
\eeq where $\mu>1$  is a constant and ${\bf  q}=(x,y)$ is the position
of one body  with respect to the other considered  fixed at the origin
of the coordinate  system, and ${\bf p}=(p_x,p_y)$ is  the momentum of
the moving particle.  The constant $\mu$ measures the  strength of the
anisotropy and  for $\mu=1$ we  recover the classical  Manev problem.
Furthermore the equation of motion can be expressed as \beq \left\{
\begin{array}{l}
\dot  {\bf  q}  =  {\bf  p}  \\  \dot{\bf  p}  =  -{\partial  H  \over
\partial{\bf q}}
\end{array}
\right. .
\label{eqmotion}
\eeq

Now consider weak anisotropies, i.e choose the parameter $\mu>1$ close
to 1. Introducing the notation $r=\sqrt{x^2+y^2}$ and $\ep=\mu-1$ with
$\ep \ll 1$ we can expand the Hamiltonian (\ref{H}) in powers of $\ep$
and obtain
  
\beq  H={1\over 2}{\bf p}^2-{1\over  r}-{b \over  r^2} +\ep  \left ({1
\over   2r}  +{b   \over  r^2}\right   )\cos^2\theta   \equiv  H_0+\ep
W(r,\theta).
\label{Hep}
\eeq 
It should  be pointed  out that  the term  $W(r,\theta)$ becomes
unbounded as $r \rightarrow 0$  so that a perturbation analysis is not
correct on  the ejection-collision orbits. This means  that the global
dynamics of the AMP cannot be completely described by perturbations to
the Manev problem even at  the limit $\ep \rightarrow 0$. However many
interesting  results  concerning the  Hamiltonian  (\ref{H}) for  weak
anisotropies (i.e. $\ep \ll 1)$  can be found studying the Hamiltonian
(\ref{Hep}), some of which are presented in this paper.

In the next section we describe  the symmetries of the AMP and we find
some properties that will be useful to find symmetric periodic orbits.
In  Section III  we prove  a  continuation theorem  for the  symmetric
periodic  orbits of ``second  kind'', i.e.  the non-circular  ones. In
Section IV we  prove a continuation theorem for  the orbits of ``first
kind'',  i.e. the  circular ones,  following the  method  developed in Ref. 8.

\section*{\large\bf II. Symmetries of the Anisotropic Manev Problem }

To  find  periodic  orbits  in  the anisotropic  problem  it  is  peculiarly
important to know the symmetries of the system, as it was, for example
observed in Refs. 5, 6. The symmetries of the problem under
discussion have been  examined in Ref. 1 and they  are the same
as  the ones  found  in Ref. 7  for  the anisotropic  Kepler
problem:
\beq \begin {array}{l} \
\hspace{3pt}      E~:~(x,y,p_x,p_y,t)\longrightarrow(x,y,p_x,p_y,t)\\\
    S_0~:~(x,y,p_x,p_y,t)\longrightarrow(x,y,-p_x,-p_y,-t)
\\\
S_1~:~(x,y,p_x,p_y,t)\longrightarrow(x,-y,-p_x,p_y,-t)\\\
S_2~:~(x,y,p_x,p_y,t)\longrightarrow(-x,y,p_x,-p_y,-t)\\\ 
S_3~:~(x,y,p_x,p_y,t)\longrightarrow(-x,-y,-p_x,-p_y,t)\\\
S_4~:~(x,y,p_x,p_y,t)\longrightarrow(-x,y,-p_x,p_y,t)\\\
S_5~:~(x,y,p_x,p_y,t)\longrightarrow(x,-y,p_x,-p_y,t)\\\
S_6~:~(x,y,p_x,p_y,t)\longrightarrow(-x,-y,p_x,p_y,-t)
\label{symmetries}
\end {array}
\eeq where $E$ is the identity.

The symmetries above can  be  interpreted   in  the   following  way: let
 $\gamma(t)$ be a solution of (\ref{eqmotion}), then
$S_i(\gamma(t))$is      another
solution for $i \in \{0,1,2,3,4,5,6\}$. For  $i\in\{0,1,2,3,4,5,6 \}$ the orbit  $\gamma(t)$ will be
called symmetric if and only if $S_i(\gamma(t))=\gamma(t)$.

Let  us  remark that  the  symmetries  in  (\ref{symmetries}), together
with the composition of functions, denoted by $\circ$, form an abelian 
group in which the operation acts according to the table below.
$$
\begin{array}{c|cccccccc}
\circ & E   & S_0 & S_1 & S_2 & S_3 & S_4 & S_5 & S_6  \\
\hline
E       & E   & S_0 & S_1 & S_2 & S_3 & S_4 & S_5 & S_6\\
S_0     & S_0 & E   & S_5 & S_4 & S_6 & S_2 & S_1 & S_3\\
S_1     & S_1 & S_5 & E   & S_3 & S_2 & S_6 & S_0 & S_4\\
S_2     & S_2 & S_4 & S_3 & E   & S_1 & S_0 & S_6 & S_5\\
S_3     & S_3 & S_6 & S_2 & S_1 & E   & S_5 & S_4 & S_0\\
S_4     & S_4 & S_2 & S_6 & S_0 & S_5 & E   & S_3 & S_1\\
S_5     & S_5 & S_1 & S_0 & S_6 & S_4 & S_3 & E   & S_2\\
S_6     & S_6 & S_3 & S_4 & S_5 & S_0 & S_1 & S_2 & E  \\
\end{array}
$$
From the table above 
it is easy to  deduce the following
\begin{proposition} 
The symmetries of the Anisotropic Manev Problem form an elementary abelian
group of order eight, i.e. a group  isomorphic to ${\bf Z}_2 \times {\bf Z}_2 \times 
{\bf Z}_2$. 
\end{proposition}  
The symmetries in (\ref{symmetries}),  (except $E$ and $S_6$) are very
useful to find  symmetric periodic orbits, especially by  means of the
continuation  method, as  we  show  in the  next  two sections.   Some
important   properties   of  the   symmetric   orbit,  summarized   in Ref. 7
, are expressed in the following lemma:

\begin{lemma}
\begin{enumerate}
\item For $i=1$ (resp. $i=2$)  we have that an orbit  $\gamma(t)$ is $S_i$-symmetric
if and only if it crosses the $x$ axis (resp. $y$ axis) orthogonally.
\item An orbit $\gamma (t)$ is $S_0$-symmetric if and only if it has a
point  on  the  zero  velocity  curve.  \item  For  $i=4,5$  an  orbit
$\gamma(t)$ is  $S_i$-symmetric if and only if  it is $S_0$-symmetric.
\item All the $S_3$-symmetric periodic orbits are periodic.
\end{enumerate}
\end{lemma}

The  properties of the  $S_i$-symmetric orbits  were first  studied by
Birkhoff\r{9} for  the restricted  three body  problem and
later  by many  other authors.  In particular  Casasayas and  Llibre\r{7} 
state a proposition  that gives a technique useful to
obtain symmetric  periodic orbits  with respect to  $S_0$, $S_1$,$S_2$
for  the anisotropic  Kepler problem  that are  verified also  for the
problem under discussion in this paper:

\begin{proposition}
\begin{enumerate}
\item  For   $i=1$ (resp. $i=2$)  we have that an orbit  $\gamma(t)$ is a
$S_i$-symmetric  periodic  orbit  if   and  only  if  it  crosses  the
$x$ axis (resp. $y$ axis)  orthogonally  at two distinct points.
\item An orbit $\gamma(t)$ is  a $S_0$-symmetric periodic orbit if and
only if it meets the zero velocity curves at two distinct points.
\item An  orbit $\gamma(t)$  is a $S_1$  and $S_2$  symmetric periodic
orbit  if  and  only  if  it  crosses  the  x  axis  and  the  y  axis
orthogonally.
\item For $i=1,2$ an orbit  $\gamma(t)$ is a $S_0$ and $S_i$-symmetric
periodic orbit  if and only  if it meets  the zero velocity  curve and
crosses the x, respectively y axis orthogonally.
\item For $i=4,5$, if an  orbit $\gamma(t)$ is $S_i$-symmetric then it
is $S_0$-symmetric and periodic.
\end{enumerate}
\end{proposition} 

Now we want  to find the symmetric periodic  orbit for the unperturbed
problem ($\ep=0$ or $\mu=1$)  and continue them to periodic solutions
of the anisotropic system (for $\ep \ll 1$).  Firstly we observe that,
by Proposition 2, the $S_i$  symmetric orbits with $i=0,4,5$ must meet
the zero  velocity curve  at two  points, i.e. there  must be  a point
where $K=0$,  but since $K$  is a constant  of motion it must  be zero
along the orbit. Therefore  such orbits are ejection-collision orbits,
are not  periodic and cannot be  studied by means  of the continuation
method. Hence  we are going  to consider the symmetric  periodic orbit
with  $i=1,2$, and  also the  ones with  $i=3$ that  are  the circular
orbits of the unperturbed problem.

To exploit  those properties  of the symmetric  periodic orbits  it is
convenient to write the equation of motion in different coordinates.

For  the $S_i$  symmetric  orbits with  $i=1,2$,  as it  was noted  in Ref. 5
, it  is convenient to  write the canonical  equations of
the restricted three body problem  using the Delaunay variables in the
rotating  frame.\r{5} Also  the Poincar\'e  synodic variables
can be used to find  symmetric periodic orbits of the restricted three
body  problem.\r{6}   The  anisotropic   Manev  problem  is
different since the Hamiltonian that describes it is time independent,
hence  the idea  of using  rotating  coordinates in  the present  case
cannot be applied. Moreover our problem is nondegenerate, however even
in our case it  is advantageous to perform a change of variables and apply a
variation  of   the  action  angle  variables   used  in Refs. 4, 12.
Here the nondegeneracy of  the problem plays a role similar
to  the  rotating  coordinate  system  in the  restricted  three  body
problem.

For the $S_3$  symmetric orbits we can instead  consider the equations
in the rotating  frame, and prove a theorem similar  to the one proved
in Ref. 8  for the  anisotropic Kepler problem  (in Ref. 8
the  author remarks that  the analysis  of the  Kepler problem  can be
redone in the Manev case, but he doesn't provide a proof) .

\section*{\large\bf III. The $\bf{S_i}$  Symmetric Orbits with $\bf{i=1,2}$}
 
We  recall  that  the  action  variables  introduced  in Refs. 4, 12 are given by

\beq \left \{
\begin{array}{l}
 I={{1 \over 2 \pi} \oint p_r  dr= - \sqrt{ K^2-2b} +{1 \over 2}{\sqrt
{2 \over |h|}}}\\ K= q_1p_2-q_2p_1
\end{array}
\right.  \eeq
where $h$ is the  energy constant and $K$ is the angular
momentum. These  variables are defined for $h<0$  and $K^2>2b$, $I>0$,
to avoid collision  orbits as well as circular  orbits. The related
frequencies are 
$$ \left \{
\begin{array}{l} \medskip
\omega_I= {1 \over (I+ \sqrt {K^2-2b})^3}\\ \medskip \omega_K={K \over
\sqrt {K^2-2b}(I+\sqrt {K^2-2b})^3},
\end{array}
\right.$$
and $\theta$ and $\phi$ are the angle variables associated  to $K$ and $I$
respectively. 

The unperturbed Hamiltonian in the new variables can
be written as
$$ H_0=-{1 \over 2(I+ \sqrt { K^2-2b})^2}.  $$

Now we can  consider new variables that are  linear combination of the
previous  ones.    They  are   defined  by  the   following  canonical
transformation

\beq \left \{
\begin{array}{l}
L=K+I\\ G=-I\\ l=\theta\\ g=\theta - \phi
\end{array}
\right .
\label{newvar}
\eeq

Where $l$ is the mean anomaly (where $l(t)= \omega_L(t-t_0)$ and $t_0$
is  the  time of  pericenter  passage), $g$  is  the longitude  of
pericenter as they are defined for the Manev problem in Ref. 13.
Moreover  also the action  variables can  be written  in terms  of the
orbital elements of the Manev problem. If we set

$$ a={1 \over 2|h|} \qquad \mbox{and} \qquad e=\sqrt{1-2(K^2-2b)|h|}
$$ 
 as   in Refs. 4, 13  then
$$   G=-a^{1/2}  \left[
1-(1-e^2)^{1/2}   \right   ]   \quad   \mbox{and}   \quad   L=-G   \pm
\sqrt{a(1-e^2)+2b} $$
where $a$ is the pseudo-semimajor axis, $e$ is
the  pseudo-eccentricity, and  the sign  + (resp.  -) holds  for $K>0$
(resp. $<0$).   The conditions to avoid collision  orbits and circular
orbits, on which  $g$ becomes meaningless, can be  written in terms of
the  orbital  elements  as  $a>0$  and $0<e<1$.  The  new  unperturbed
Hamiltonian is  \beq H_0=-{1 \over 2(-G+ \sqrt  { (G+L)^2-2b})^2} \eeq
so   the   perturbed  equations   of   motion   become  \beq   \left\{
\begin{array}{l} \dot{L}  =-{\partial (H_0+\ep W) \over  \partial l} =
-\ep{\partial  W \over  \partial  l} \medskip  \\ \dot{G}=  -{\partial
(H_0+\ep  W)  \over  \partial  g}=-\ep{\partial W  \over  \partial  g}
\medskip\\   \dot{l}=  {\partial  (H_0+\ep   W)  \over   \partial  L}=
\omega_L+\ep{\partial  W  \over   \partial  L}  \medskip  \\  \dot{g}=
{\partial (H_0+\ep W) \over \partial G}= \omega_G+\ep{\partial W \over
\partial G}
\end{array}
\label{pe}
\right .   \eeq where $W$ is  expressed in the new  variables and
$$
\left \{
\begin{array}{l}
\omega_L=\omega_K={G+L                                            \over
(-G+\sqrt{(G+L)^2-2b})^3\sqrt{(G+L)^2-2b}}                   \medskip\\
\omega_G=\omega_K-\omega_I={G+L-\sqrt{(G+L)^2-2b}                 \over
(-G+\sqrt{(G+L)^2-2b})^3 \sqrt{(G+L)^2-2b}}
\end{array}
\right. $$

With  these preparations, i.e. the introduction  of the  action angle
variables (\ref{newvar}), we are well on our way to estabilishing the 
following result:

\begin{theorem} Let $\gamma(t)$ be an  $S_i$-symmetric periodic orbit  of the Manev problem with $i=1,2$.
 Let the period  be $\tau$ and set $\ep=\mu-1$ with  $\ep \ll 1$. Then
 there  exists a  $\tau$-periodic  solution of  the Anisotropic  Manev
 problem            $\gamma_\epsilon(t)$           such           that
 $\gamma_\epsilon(t)=\gamma(t)+O(\epsilon)$.
\end{theorem} 
PROOF: Let's consider  an $S_1$ symmetric orbit of  period $\tau= 2\pi
m/k$ ($m$, $k$ relatively prime  integers).  We remark that, since the
equations  of  motion are  autonomous,  we  can  reduce to  study  the
symmetric orbits that  have either the pericenter or  the apocenter on
the positive $x$ axis at $t=0$.

If at $t=0$, $\ep=0$, the pericenter  of this orbit is on the positive
$x$ axis,  and it  is crossing the  $x$ axis perpendicularly,  we have
\beq g(0)=0  \quad \mbox{and} \quad  l(0)=0.  \eeq Since  the periodic
orbit is $S_1$ symmetric, by Proposition 2, at the half period one has
\beq g(\tau/2)=m\pi \qquad l(\tau/2)=k\pi
\label{periodicity}
\eeq that follows from the solution of (\ref{pe}) for $\ep=0$: \beq
\begin{array}{l}
L=\mbox{const.} \qquad \qquad G=\mbox{const.}  \\ l= \omega_L t \qquad
\qquad \quad ~ g=\omega_G t
\end{array}
\label{periodicity1}
\eeq  Now  if, for  $\ep  \neq 0$  we  consider  only $S_1$  symmetric
solutions of (\ref{pe}), it follows from the implicit function theorem
that if the functional determinant \beq D=det \left (
\begin{array}{cc}
\partial  l /  \partial L  &  \partial l/  \partial G  \\ \partial  g/
\partial L & \partial g/ \partial G
\end{array}
\right  )  \neq  0  \eeq  at  \beq t=\tau/2  \qquad  \ep=0  \eeq  then
(\ref{periodicity}) would  be satisfied for  $\ep>0$ . To  compute the
determinant we can  by analyticity substitute (\ref{periodicity}) into
(\ref{periodicity1})  to find  out at  the time  $t=\tau/2$  that \beq
D={6b(\tau/2)^2 \over (-G+\sqrt{(G+L)^2-2b})^7((G+L)^2-2b)^{3/2}} \neq
0
\label{det}
\eeq Thus the  existence of $S_1$ symmetric periodic  orbits of period
$\tau$ obtained from the  $\tau$ periodic $S_1$ symmetric solutions of
the  unperturbed problem,  that at  $t=0$ have  the pericenter  on the
positive $x$ axis, is readily established.

On  the other  hand, if  at $t=0$,  $\ep=0$, the  apocenter is  on the
positive $x$ axis, and it is crossing the $x$ axis perpendicularly, we
have \beq  g(0)=\pi/ \lambda \quad  \mbox{and} \quad l(0)=-\pi/\lambda
\eeq where  $\lambda= (\omega_L-\omega_G)/ \omega_L$.   By Proposition
1, at  the half period we have  \beq g(\tau/2)=(m+1/\lambda)\pi \qquad
l(\tau/2)=(-1/\lambda+k)\pi.
\label{halfp}
\eeq Instead of computing the functional determinant directly, in this
case,  it  is  easier to  consider  the  new  variables given  by  the
relations, \beq \left\{
\begin{array}{l}
\tilde  L=L\\  \tilde  G=G\\  \tilde l=  l+\pi/\lambda_0\\  \tilde  g=
g-\pi/\lambda_0
\end{array}
\right.
\label{canon}
\eeq that define a family of canonical transformations parametrized by
$\lambda_0(L_0,G_0)$. For each orbit choose a different transformation
from the  family (\ref{canon}),  where $\lambda_0=\lambda$ is  a fixed
quantity  defined by  the  value  of the  action  variables along  the
periodic orbit under consideration.

The equations  (\ref{halfp}), expressed in  the new variables,  are of
the  same   form  as  in  (\ref{periodicity}).   Thus  the  functional
determinant, in the  new variables, is exactly $D$,  and the existence
of the remaining $S_1$-symmetric $\tau$-periodic orbits follows.

Now the  proof for  the $S_2$-symmetric orbits  can be done  along the
same lines.  Consider an
$S_2$ symmetric periodic orbit of period $\tau=2\pi m/k$. If at $t=0$,
$\ep=0$ the pericenter of the orbit is on the positive $y$ axis and it
is  crossing the  $y$ axis  perpendicularly, we  have  \beq g(0)=\pi/2
\quad \mbox{and} \quad  l(0)=0 \eeq Since the periodic  orbit is $S_2$
symmetric  one has,  at  the half  period  \beq g(\tau/2)=m\pi  +\pi/2
\qquad \qquad l(\tau/2)=k \pi.
\label{periodicity1a}
\eeq Now we consider only  $S_2$ symmetric solutions of (\ref{pe}) for
$\ep \neq 0$ again it  follows from the implicit function theorem that
if the determinant $D$ computed  at $t=\tau/2$ for $\ep=0$ is non zero
then  (\ref{periodicity1a}) would  be  satisfied for  $\ep>0$.  It  is
trivial to  see from (\ref{det}) that  $D \neq 0$, and  hence we found
$S_2$ symmetric  periodic orbits for  the perturbed problem. 

For the $S_2$-symmetric orbits  having the apocenter  on the  positive
$x$ axis at $t=0$ the canonical transformation (\ref{canon}) can be used.
Again we find the same expression  for the functional determinant and
hence, by  the implicit  function  theorem, the  existence  of the 
remaining $S_2$-symmetric periodic orbits is proved.

It  is interesting  to remark that  Theorem 1 and its proof can be 
easly extended to consider any $S_i$-symmetric perturbation with $i=1,2$
and a very general class of nondegenerate integrable Hamiltonians,
however  such a generalization is trivial and not strictly related to
the problem under consideration and hence it will not be discussed
any further.

We can also observe that   for $b=0$,
i.e.  for  the  Kepler  problem,  the determinant  in  (\ref{det})  is
zero.  Thus  in  the  case  of the  Anisotropic  Kepler  Problem,  the
continuation theorem proved above cannot be applied, and the existence
of   symmetric  periodic   orbits   of  ``second   kind''  (for   weak
anisotropies)  remains unclear.  On  the other  hand the  continuation
theorem that  we prove in the  next section (for  the circular orbits)
can  be applied  to the  Anisotropic Kepler  Problem\r{8} and
hence at least the existence of symmetric periodic orbits of the first
kind is a well established fact.
\section*{\large\bf IV. The $\bf{S_3}$ Symmetric Orbits}

Again  we can  consider  the
 Anisotropic  Manev  Problem taking  the
parameter  $\mu$ close to  1. Let  the flow  $\Phi(t,({\bf r},\dot{\bf
r}),\mu)$ of the equation of motion  (1). In this section we prove the
following theorem:

\begin{theorem}  
Let ${\bf  r}^0(t)$ be a  $S_3$-symmetric periodic orbit of  the Manev
problem, i.e. a circular one. Set  $\ep=\mu -1$, and let $\tau$ be the
period of ${\bf r}^0(t)$. Then there exists a $\tau$-periodic solution
$  \Phi (t,({\bf r}(\ep),\dot{\bf  r}(\ep)),\ep)$ of  the Anisotropic
Manev problem such  that $\Phi(t,({\bf r}(0),\dot{\bf r}(0)),0)=({\bf
r}^0(t),\dot{\bf r}^0(t))$.
\end{theorem}
\subsection*{\normalsize\bf A. The equation of motion}
Now using the same notation as in Ref. 8 let $ \bf{r}^0(t)$ be a
circular solution of the Manev  problem which correspond to $\mu=1$ in
the $xy$-plane,  $\omega$ its  angular speed and  $a$ its  radius. For
$\ep=\mu-1\neq 0$ we set, \beq {\bf r}(t,\ep)={\bf r}^0(t)+\ep~{\bf s}
(t,\ep)
\label{circular}
 \eeq Expanding $\nabla  H$ in powers of $\mu  -1$ sufficiently small,
after substituting the expression for $\bf r$ given above, considering
the notation ${\bf r}^0(t)=x_0(t)+iy_0(t)$ and ${\bf s} =u+iv$ we have
that ${\bf r}(t,\ep)$  is a solution of equation  of motion defined by
(\ref{H})  if, and  only if,  ${\bf s}(t,\ep)$  is a  solution  of the
equations

\beq
\begin{array}{l} \ddot u=- \left({1 \over a^3}-{3x_0^2 \over a^5}-{8bx_0^2 \over a^6}+{2b \over a^4} \right)u +\left({3x_0y_0 \over
a^5}+{8bx_0y_0 \over a^6}\right)v +\eta(t)+O(\ep) \\\noalign{\medskip}
\ddot v=\left({3x_0y_0  \over a^5}+{8bx_0y_0 \ov a^6}\right)u-\left({1
\ov    a^3}-{8by_0^2    \ov     a^6}-{3y_0^2    \ov    a^5}+{2b    \ov
a^4}\right)v+\xi(t)+ O(\ep)
\end{array} 
\label{eqofmotion}
\eeq 
where 
$$
\begin{array}{l}
{\eta(t)}={3x_0y_0^2  \over a^5}  +{4bx_0y_0^2  \over a^6}  \medskip\\
{\xi(t)}={3y_0^2\over 2a^5}-{y_0 \over a^3}+{4by_0^3 \over a^6}-{2by_0
\over a^4}
\end{array}
$$
Consider the orthonormal frame in ${\bf R}^2$,$~{\bf e}_1(t)$ and
${\bf e}_2(t)$ defined by

$${\bf e}_1={{\bf r}^0 \over |{\bf r}^0|}=e^{i \omega t}=\cos \omega t
+i \sin  \omega t,  \qq {\bf e}_2=  i{\bf e}_1  $$ and using  the same
notation as  in Ref. 8  where
$$ {\bf  s}= x_1 {\bf  e}_1 +x_2
{\bf e}_2, \qquad \dot{\bf  s}=y_1{\bf e}_1+y_2{\bf e}_2 $$ equation
(\ref{eqofmotion}) can be written in  an equivalent form as: \beq {\bf
\dot z}=A_0(t)+ A{\bf z} + O(\ep),
\label{rotsystem1}
\eeq
 where $ {\bf z}=(x_1,x_2,y_1,y_2)^T$, and

$$ A_0=\left (\
\begin{array}{c} 
0\\ 0\\ \alpha(t)\\ \beta(t)
\end{array}\right)
\qquad             A=\left            (\begin            {array}{cccc}
0&\omega&1&0\\\noalign{\medskip}-\omega&0&0&1
\\\noalign{\medskip}2\,{\omega}^{2}+2\,{\frac {b}{{a}^{4}}}&0&0&\omega
\\\noalign{\medskip}0&-{\omega}^{2}&-\omega&0\end {array}\right ) $$
where 
$$
\begin{array}{l}
\alpha(t)\cos\omega          t-\beta(t)\sin\omega          t=\eta(t)\\
\alpha(t)\sin\omega t +\beta(t)\cos\omega t=\xi(t)
\end{array}
$$ or equivalently, \beq
\begin{array}{l}
\alpha(t)=\sin^2\omega     t\left     ({1\over     2a^2}+{2b     \over
a^3}\right)\medskip\\  \beta(t)=-\sin\omega t~  \cos\omega  t \left({1
\over a^2}+{2b \over a^3}\right)
\label{beta}
\end{array}
\eeq The eigenvalues of $A$ are $0$, with multiplicity two, $ {i \over
a^{3/2}}$  and $-{i  \over  a^{3\  2}}$. One  of  the two  eigenvalues
vanishes because the  system is autonomous, and the  second due to the
presence of the first integral $H$.

Now  consider the  real Jordan  form  $J$ of  $A$. The  matrix $J$  is
defined by the relation $ J= {\cal  T} ^{-1} A {\cal T} $ where ${\cal
T}$ is

$$\scriptstyle{\mathrm{
{\cal  T}= \left  (\begin{array}{cccc} \scriptstyle{2\,{\omega}^{2}{a}^{3}}&0&{\frac
{{\omega}^{2}{a}^{4}          +2\,b}{a}}&0\\\noalign{\medskip}0&-{\frac
{\omega\left  (3  \,{\omega}^{2}{a}^{4}+2\,b\right )}{a}}&0&-2\,{\frac
{\omega{a   }^{2}\left  ({\omega}^{2}{a}^{4}+2\,b\right   )}{\left(  a
\right )^{3/2}}}\\\noalign{\medskip}0&{1  \over 2}\,{\frac {4\, a\left
({\omega}^{2}{a}^{4}+b\right        )+2\,\left        ({\omega}^{2}{a}
^{4}+2\,b\right        )^{2}}{{a}^{5}}}&0&{        \frac        {\left
({\omega}^{2}{a}^{4}+2\,b\right
)^{2}}{{a}^{7/2}}}\\\noalign{\medskip}-{\frac              {\omega\left
({\omega}^{2}    {a}^{4}+2\,b\right    )}{a}}&0&-{\frac   {\omega\left
({\omega}^{2} {a}^{4}+2\,b\right )}{a}}&0\end{array}\right )}}$$
 and the  columns of  ${\cal T}$ are  the generalized  eigenvectors of
$A$.

The  vector  $J_0={\cal  T}^{-1}A_0$  and  the matrix  $J$  are:  $$
J_0=\left( \begin{array}{c} j_1(t) \\ j_2(t) \\ j_3(t) \\ j_4(t)
\end{array} \right ), \qquad
J=\left   (\begin   {array}{cccc}   0&0&0&0\\\noalign{\medskip}1&0&0&0
\\\noalign{\medskip}0&0&0&{\frac         {\sqrt         {a}}{{a}^{2}}}
\\\noalign{\medskip}0&0&-{\frac  {\sqrt  {a}}{{a}^{2}}}&0\end  {array}
\right ) $$

where the fact that $j_1(t)=(2\omega^3a^2-{\omega(\omega^2a^4+2b)\over
a^2})^{-1}\beta(t)$ is  the only information about $J_0$  that we need
to retain.  Furthermore we  remark that $\omega^2a^4-a-2b=0$ gives the
relation  between $a$ and  $\omega$ and  solving this  equations gives
only one positive solution (for $b>0$).

Letting ${\bf z}={\cal T}\zeta$, the equation of motion becomes

\beq \dot\zeta=J_0(t)+J\zeta +O(\ep),
\label{rotsystem2}
\eeq

and its flow is given by

\beq \psi(t,\zeta, \ep)=\gamma(t)+ e^{Jt}+O(\ep)
\label{jacobflow}
\eeq

where by  the variation of  constants \beq \gamma(t)=  e^{Jt} \int_0^t
e^{-Js} J_0(s) ds
\label{variation}
\eeq

Therefore we have

$$          e^{Jt}=\left(\begin                 {array}{cccc}
 1&0&0&0\\\noalign{\medskip}t&1&0&0        \\\noalign{\medskip}0&0&\cos
 {\frac  {\sqrt  {a}}{{a}^{2}}}t&\sin  {\frac  {\sqrt  {a}}{{a}^{2}}}t
 \\\noalign{\medskip}0&0&-\sin  {\frac  {\sqrt  {a}}{{a}^{2}}}t  &\cos
 {\frac {\sqrt {a}}{{a}^{2}}}t \end {array} \right ) $$

and from (\ref{variation}) we obtain
\beq \gamma(t)=\left (
\begin{array}{c}
\gamma_1(t)\medskip\\    \gamma_2(t)\medskip\\   \gamma_3(t)\medskip\\
\gamma_4(t)
\end{array} \right )
\eeq
where    we   retain    only   the   information    that
\beq
\gamma_1(t)=(2\omega^3a^2-{\omega(\omega^2a^4+2b)\over
a^2})^{-1}\int_0^t\beta(s)ds.
\label{gamma1}
\eeq
\subsection*{\normalsize\bf B. The periodicity equation}

Since  the  Hamiltonian  $H$  of  the  anisotropic  Manev  problem  is
$S_3$-symmetric,  as  we have  shown,  we  can  write the  periodicity
equation as in\r{8}, \beq \Phi \left({\tau \over 2},({\bf r},
{\bf \dot r}),\ep \right)=-({\bf r}, {\bf \dot r}).
\label{periodicity3}
\eeq Then it easy to  check that $\Phi(t,({\bf r}, {\bf \dot r}),\ep)$
is a periodic  solution of the equation of  motion with period $\tau$.
To find periodic solutions we have to verify that (\ref{periodicity3})
is   satisfied  for   a  family   of  initial   conditions.   Equation
(\ref{periodicity3}) in $\zeta$ coordinates  is \beq \psi \left( {\tau
\over 2}, \zeta , \ep \right) -\zeta=0 \eeq where $\psi(t,\zeta, \ep)$
is  the   flow  of  (\ref{rotsystem2}).   Let  us  denote   by  ${\cal
P}(\zeta,\ep)$  the  left  hand   side  of  the  periodicity  equation
(\ref{periodicity3}),  that  is,  let  \beq  {\cal  P}(\zeta,\ep)=\psi
\left({\tau/2},                    \zeta,\ep                   \right)
-\zeta=\gamma(\tau/2)+\left(e^{J{\tau\over 2}}-I\right)= 0.
\label{p}
\eeq

Using (\ref{jacobflow})  we notice that the requirement  \beq {\cal P}
(\zeta^\ast,0)=\gamma(\tau/2)+\left(     e^{J{\tau     \over     2}}-I
\right)\zeta^\ast=0,    \eeq    imposes    the    restrictions    \beq
\gamma_1(\tau/2)=0  \quad  \zeta_1^\ast=-{2\over \tau}\gamma_2(\tau/2)
\quad \zeta_2^\ast=\mbox{arbitrary}\quad \eeq and \beq
\begin{array}{l}
\zeta_3^\ast={1                                                   \over
2(1-\cos\alpha^\ast)}\left(-\gamma_3(\tau/2)(\cos\alpha^\ast
-1)+\gamma_4(\tau/2)\sin\alpha^\ast \right)\medskip\\

\zeta_4^\ast={-1                                                  \over
2(1-\cos\alpha^\ast)}(\gamma_3(\tau/2)\sin\alpha^\ast+\gamma_4(\tau/2)(\cos\alpha^\ast-1))
\end{array}
\eeq  where $\alpha^\ast=\pi  (1+2b/a)^{-1/2}$.  It  easy to  see from
(\ref{beta}) and  (\ref{gamma1}) that $\gamma_1(\tau/2)=0$, therefore,
we                                                                 take
\beq\zeta^\ast=(\zeta_1^\ast,\zeta_2^\ast,\zeta_3^\ast,\zeta_4^\ast)^T,\eeq
with  $\zeta_2^\ast$ arbitrary,  for the  moment. Now  using  the flow
(\ref{jacobflow}) , we  determine that the Jacobian matrix  of $\cal P$ with
respect   to   the   variables   $\zeta$  evaluated   at   the   point
$(\zeta^\ast,0)$  is given  by \beq  \left (  \begin{array}{cccc}\ 0&0
&0&0\\\ \tau/2&0&0&0  \\\ 0&0&  \cos \alpha^\ast -1&  \sin \alpha^\ast
\\\ 0&0&-\sin  \alpha^\ast & \cos \alpha^\ast  -1\end{array} \right ).
\eeq

Consider the  system of three  equations formed by those  in (\ref{p})
corresponding   to  the   indices   i=2,3,4  and   fix  the   variable
$\zeta_2=\zeta_2^\ast$.  Its  Jacobian matrix has  determinant $\tau (
1-\cos\alpha^\ast)$,   that   is  always   positive   since  $0<   \pi
(1+2b/a)^{-1/2}\leq  \pi $.  Therefore  the implicit  function theorem
guarantees the existence  of analytic functions $\zeta_i=\zeta_i(\ep),
~i=1,3,4$  in  a neighborhood  of  $\ep=0$,  satisfying the  equations
\beq{\cal     P}_i(\zeta,    \ep)=0,     \q     (i=2,3,4)\eeq    where
\beq\zeta(\ep)=(\zeta_1(\ep),
\zeta^\ast_2,\zeta_3(\ep),\zeta_4(\ep))\eeq   and   such   that   \beq
\zeta_i(0)=\zeta_i^\ast \q  (i=1,2,3,4).  \eeq It remains  to show, in
order to have periodicity,  that also the remaining equation \beq {\cal
P}_1(\zeta(\ep),\nu(\ep),\ep)=0,  \eeq  is  satisfied in  a,  possibly
smaller  neighborhood of  $\ep=0$.  That  will be  done employing  a first
integral of the system under discussion, i.e. the Hamiltonian.

\subsection*{\normalsize\bf C. Integral of motion}
Since the  Hamiltonian is  a integral of  motion of the  problem under
discussion we can apply the same analysis as in Refs. 8, 14. In
particular using  the same notations  as in Ref. 8 we can define
$$ H_\ep({\bf z},t)=H({\bf r},\dot{\bf r},\ep),$$
where $H_\ep({\bf z},t)$ is a time-dependent, $\tau$-periodic first integral 
for system (\ref{rotsystem1}). The above integral satisfies the following 
relation
\beq
H_\ep({\bf z},t+\tau/2)=H_\ep({\bf z},t) 
\label{int1}
\eeq
for all $t$, since $H(-{\bf r},-\dot{\bf r})=H({\bf r},\dot{\bf r})$,
 $~{\bf r}(t)={\bf r}^0(t)+\ep~{\bf s}(t)$ and 

$${\bf r}^0(t+\tau/2)=-{\bf r}^0(t)\q,\q {\bf s}({\bf z},t+\tau/2)=-{\bf s}({\bf z},t).$$

Performing a change of coordinates we can define ${\cal H}_\ep(\zeta,t)=H_\ep({\cal T}\zeta,t)$,
hence (\ref{int1}) can be written as
\beq
{\cal H}_\ep(\zeta, t+\tau/2)={\cal H}_\ep(\zeta,t).
\label{int2}
\eeq
Moreover since ${\cal H}_\ep$ is an integral of motion it verifies that
\beq 
{\cal H}_\ep(\phi(\zeta, \ep,t))=H_\ep(\zeta,0).
\label{int3}
\eeq
Thus applying equations (\ref{int2}-\ref{int3}) it follows that
$${\cal H}_\ep(\psi(\tau/2,\zeta,\ep),0)={\cal H}_\ep(\zeta,0)$$
and by means of the Mean Value Theorem we obtain
\beq  \nabla   _\zeta  {\cal  H}_\ep   (\tilde{\zeta},0)  \cdot  {\cal
P}(\zeta,\ep)=0,
\label{meanval}
\eeq

where$\nabla_\zeta{\cal H}_\ep$ is the gradient of ${\cal H}_\ep$ with
respect  to $\zeta$,  and $\tilde{\zeta}$  is a  point on  the segment
joining $\zeta$ to $\psi(\tau/2,\zeta,\ep)$.

Expanding $\Psi(\ep)=\psi(\tau/2,\zeta,\ep)$ in power of $\ep$ sufficently small 
it is easy to show (see Ref. 8)  that $\Psi(\ep)=\zeta^\ast+O(\ep)$ and consequently 
$$\tilde{\zeta}=s\zeta(\ep)+(1-s)\Psi(\ep)=\zeta^\ast +O(\ep)$$
for some $s\in (0,1)$. 
Moreover if we also expand the Hamiltonian $H_\ep({\bf z},0)$ in powers of $\ep$ we get
$$H_\ep({\bf z},0)=H_0+\ep(H_1+H_2\cdot {\bf z} )+O(\ep^2)$$
or, in $\zeta$ coordinates  
\beq  {\cal   H}_\ep(\zeta,0)={\cal  H}_0  +   \ep({\cal  H}_1  +{\cal
H}_2\cdot \zeta) +O(\ep^2), 
\eeq

where  ${\cal  H}_0=H_0=\left(  {1\over 2}\omega^2a^2-{1\over  a}  -{b
\over  a^2  }\right)$, ${\cal  H}_1=H_1$  and  ${\cal H}_2={\cal  T}^T
H_2={\cal T}^T(a^{-2}+2ba^{-3},0,0,a\omega)=(  a \omega^2 \zeta_1,0,0,0)$.
Hence   we  obtain 
 \beq   {1  \over  \ep}   \nabla_\zeta  {\cal
H}_\ep(\tilde{\zeta},0)={\cal  H}_2 +O(\ep).  
\eeq 
With these preparations   equation (\ref{meanval}) reduces  to the equation 
in the unknown  ${\cal P}_1$
\beq [a \omega^2+O(\ep)]{\cal P}_1=0.   \eeq 
since, for  small $\ep$,  we  already   found in Sec. IV B that  ${\cal  P}_i=0$ for $i=2,3,4$.

It is easy to see that for $\ep=0$ the  equation above   has   solution  ${\cal
P}_1=0$.  Thus,   by  continuity,  $[a   \omega^2+O(\ep)]$  is
different from zero for $\ep$  sufficiently small. Therefore for such values of
$\ep$  this equation has a unique solution that is the trivial
one. Consequently  the  remaining  equation   $${\cal
P}_1(\zeta(\ep),\ep)=0,$$  is  also   satisfied in a possibly smaller neighborhood of $\ep=0$.
Hence  all  the equations  of  the  periodicity  system (\ref{p})  are  satisfied  when  
$\zeta=\zeta(\ep)$, as long as $\ep$ is sufficiently small. This  completes 
the proof of Theorem 2.

\qq\qq

\noindent
\section*{\large\bf Acknowledgments}
 
The author is grateful to Professor Florin Diacu for  
his enlightening comments and suggestions. This work  was supported 
by an University of Victoria Fellowship and  a Howard E. Petch Research
Scholarship.

\qq\qq

\section*{\large\bf References}

\r{1} S. Craig, F. Diacu, E. A. Lacomba, E. Perez, ``On the
anisotropic  Manev  problem'', J.  Math.  Phys.,  {\bf 40},  1359-1375
(1999)

\r{2} A. Einstein, ``Zum Quantensatz von Sommerfeld und
 Epstein'', Verhandl. Deutsch. Phys. Gesellsch., {\bf 19}, 82-92 (1917)

\r{3}  Y. Hagiara, {\it  Celestial Mechanics}  (MIT Press,
Cambridge, MA, 1975), Vol. II part I

\r{4}  F. Diacu, M.  Santoprete,  ``Nonintegrability  and
Chaos in  the Anisotropic Manev problem'', Physica  D, {\bf 156},39-52
(2001)


\r{5}  R.B. Barrar, ``Existence  of periodic  orbits  of the
second kind in the restricted problem of three bodies'' , Astronomical
J. {\bf 70},3-4 (1965)

\r{6} A.  Milani, ``Stability and  Bifurcation of Symmetric
Periodic Orbits'', Rend. Circ. Mat. Palermo {\bf 34}, 161-191 (1985)

\r{7} J. Casasayas, J. Llibre, ``Qualitative Analysis of
the Anisotropic Kepler Problem'', Mem.  Am. Math. Soc. {\bf 52}, 1-115
(1984)

\r{8}  C.   Vidal,  ``Periodic  Solutions   for  Any  Planar
Symmetric Perturbation of the Kepler Problem'', Cel. Mech., Dyn. Astr,
119-132, (80) (2001)

\r{9}  G.  Birkhoff, ``The  Restricted  Problem of  Three
Bodies'', Rend. Circ. Mat. Palermo, {\bf 39}, 1-70 (1915)

\r{10}  H.  Poincar\'e,  ``Les  Methods  Nouvelles  de  la
Mecanique Celeste'', {\it Gauthier-Villars} (1892)

\r{11}  V. Szebehely,  {\it Theory  of  Orbits} (Academic
Press, New York, 1967)

\r{12} W. Thirring, {\it  A Course in Mathematical Physics
I, Classical Dynamical Systems} (Springer Verlag, New York, 1978)

\r{13} C.  Stoica, `` A  proper set of orbital  elements and
perturbation equations in Maneff-type fields'', preprint

\r{14} C.  L. Siegel, Moser J.K., {\it  Lectures on Celestial
Mechanics} ( Springer-Verlag, Berlin, 1971)

\end{document}